\documentclass[12pt,a4paper]{article}
\usepackage[textheight=23cm,textwidth=15.5cm]{geometry}

\usepackage[utf8x]{inputenc}
\usepackage{graphicx}
\usepackage{cite}
\usepackage{here}
\usepackage{amsmath}
\usepackage{amssymb}

\title{Automated Selection of Active Orbital Spaces}
\author{Christopher J. Stein and Markus Reiher\thanks{corresponding author: markus.reiher@phys.chem.ethz.ch; Phone: +41446334308; Fax: +41446331594}}

\begin{document}

\maketitle

\hspace*{-0.9cm}

\begin{center}
ETH Z\"urich, Laboratory of Physical Chemistry, \\ Vladimir-Prelog-Weg 2, 8093 Z\"urich, Switzerland. \\
\end{center}

\hspace*{0.5cm}

\begin{abstract}
One of the key challenges of quantum-chemical multi-configuration methods is the necessity to manually select orbitals for the active space.
This selection requires both expertise and experience and can therefore impose severe limitations on the applicability of this most general class of \textit{ab initio} methods.
A poor choice of the active orbital space may yield even qualitatively wrong results.
This is obviously a severe problem, especially for wave function methods that are designed to be systematically improvable.
Here, we show how the iterative nature of the density matrix renormalization group combined with its capability to include up to about one hundred orbitals in the active space can be exploited for a systematic assessment and selection of active orbitals.
These benefits allow us to implement an automated approach for active orbital space selection, which can turn multi-configuration models into black box approaches.
\end{abstract}

\newpage
\section{Introduction}

If a molecule features partially occupied close-lying frontier orbitals, single-configuration methods such as Hartree--Fock (HF) --- and also contemporary approaches based on single-determinant Kohn--Sham density functional theory (DFT) --- will not provide a reliable approximation of the electronic wave function.
This can only be achieved by a superposition of configurations (configuration interaction, CI).
Since the number of possible configurations is in general enormous, a selection of configurations becomes unavoidable in practice.
However, a configuration selection procedure should lead to a well-defined approximation of an electronic state in order to avoid any bias or arbitrariness in the wave function model.
Complete active space (CAS) approaches\cite{ruedenberg1982,roos1980,werner1985,knowles1985} are well-defined models, because \textit{all} possible configurations within some orbital subspace are considered.
In general, an active space as compact as possible is desirable that includes all statically correlated orbitals to clearly discriminate between static and dynamic correlation.\cite{pulay2015}

The restriction of the size of the orbital space then requires a recipe for the selection of the active orbitals, causing the construction of the CAS to feature an element of ambiguity.
As a consequence, extensive studies were performed to deduce general selection rules.\cite{roos1987,pierloot2001,pierloot2003,veryazov2011}
However, the CAS selection still remains a delicate task that was described as ''a tremendous challenge''\cite{head-gordon2006}.
This manual selection of orbitals was also rated as ''highly subjective and can lead to serious problems''\cite{bach2006} and it was realized 
that one ''must experiment''.\cite{gagliardi2008}

The recent developments in the density matrix renormalization group (DMRG)\cite{white1992,white1993} applied to quantum chemistry\cite{legeza2008,chan2008,chan2009,reiher2010,chan2011,marti2011,keller2014,kurashige2014,wouters2014a,yanai2015,szalay2015,knecht2015} offer a remedy for these disadvantages for two reasons:
\begin{enumerate}
\item With DMRG one can perform calculations for active orbital spaces of up to about one hundred orbitals as opposed to a rather limited size of the CAS in traditional methods due to exponential scaling (restricted to about 18 electrons in 18 orbitals\cite{molcas8}).
\item As an iterative method, DMRG optimizations can produce a qualitatively correct approximate wave function after comparatively few iterations before energy convergence is reached.\cite{moritz2007}
\end{enumerate}
Since DMRG is rather flexible with respect to the size of the CAS and partially converged DMRG solutions can be obtained quickly, a protocol for the automated active orbital space selection is in reach.
Such a protocol will turn CAS approaches into black box methods.
Different active orbital spaces can be automatically assessed before a final calculation is driven to energy convergence.
Scoring functions will be required to supply orbital selection criteria.
For this purpose, natural orbital occupation numbers\cite{jensen1988,pulay1989,pulay2015,wouters2014,krausbeck2014} and entropy-based entanglement measures\cite{legeza2003,rissler2006,boguslawski2012} defined with respect to individual orbitals can be employed.
In this work, we show how a DMRG-based automated active orbital space selection can be accomplished.

\section{Computational Methodology}
\label{comp_sect}

All structures of molecules considered for this work are listed in Table \ref{comp_met}. Those, that were not available in the literature were optimized with \textsc{turbomole} version 6.5\cite{ahlrichs1989}  with DFT functionals and basis sets defined in the same table.
Four different types of active orbitals, namely HF, split-localized, CAS self-consistent field (CASSCF) orbitals from a small CAS and DMRG-SCF orbitals, were chosen for the DMRG calculations. 
Split-localized orbitals\cite{chan2015} were obtained by separately localizing the occupied and virtual orbitals of a preceding HF calculation with the Pipek-Mezey method.\cite{pipek1989}
Partially converged DMRG-SCF orbitals (from an exploratory large-CAS DMRG calculation) 
were obtained from calculations with loose convergence thresholds and a low number of renormalized block states $m$.
The active orbitals for the small-CAS CASSCF calculations and the DMRG-SCF orbitals are specified where required.
All orbitals were generated with \textsc{Molcas}\cite{molcaspaper,molcas8} and with our matrix product operator based DMRG program \textsc{QCMaquis}, respectively.\cite{qcmaquis}
The DMRG calculations and the evaluation of entanglement measures were performed with \textsc{QCMaquis}.\cite{qcmaquis}\\

\begin{table}
\caption{\small Overview of density functionals and basis sets applied for structure optimization and orbital generation.}
\begin{tabular}{llll}
\hline \hline
molecule & method & \multicolumn{2}{l}{\hspace{1.3cm}basis sets for:}\\
 &  & \hspace{0.1cm}structure & orbitals \\
\hline
MnO$_4^-$ & B3LYP\cite{parr1988,frisch1994} & \hspace{0.1cm}def2-TZVP\cite{weigend1998} &ANO-S\cite{pierloot1995} \\
chloroiron corrole & B3LYP\cite{parr1988,frisch1994} & \hspace{0.1cm}def2-TZVP\cite{weigend1998} & ANO-S\cite{pierloot1995} \\
CrF$_6$/CrF$_6^{3-}$ & BP86\cite{becke1988,perdew1986} & \hspace{0.1cm}def2-TZVP\cite{weigend1998} & F: ANO-S\cite{pierloot1995}, Cr: Stuttgart ECP$^a$\cite{dolg1987} \\
oxo-Mn(salen) &Ref. \citenum{ivanic2004} & \hspace{0.1cm}- &cc-pVDZ\cite{dunning1989,peterson2005} \\
Cu$_2$O$_2^{2+}$ & Ref. \citenum{cramer2006} & \hspace{0.1cm}- & O: ANO-S\cite{pierloot1995}, Cu: Stuttgart ECP$^a$\cite{dolg1987}\\
\hline
\hline
\multicolumn{4}{l}{$^a$denotes an effective core potential and associated basis functions.}
\end{tabular}
\label{comp_met}
\end{table}

We introduce the notation 'DMRG[$m$]($N$,$L$)\#\textit{orbital\_basis}', to denote the number of renormalized block states $m$, the number of electrons $N$, and orbitals in the CAS  $L$, and the  orbital basis of the DMRG calculation.
Whenever an orbital basis is specified by the \#-notation, this implies a DMRG-CI calculation without further orbital optimization.
The active space in the CASSCF calculations is specified as CAS($N$,$L$)-SCF.

The error in the DMRG wave function introduced through the restriction of the number of block states $m$ can be quantified by the truncation error $\epsilon$\cite{hess2003},
\begin{equation}
\label{truncerr}
\epsilon = \Vert |\Psi \rangle - |\tilde{\Psi} \rangle \Vert^2 = 1- \sum_{\alpha = 1}^m w_\alpha ,
\end{equation}
where $\Psi$ denotes the target wave function, $\tilde{\Psi}$ the approximated wave function, and $w_\alpha$ the eigenvalues of the reduced density matrix of the active subsystem.
An extrapolation of the DMRG energies $E(m)$ to a truncation error of zero --- and therefore to the target energy $E_\mathrm{Extrapol.}$--- is possible based on results obtained with varying $m$.\cite{chan2002,hess2003,marti2010} 
We apply a linear fit function with slope $a$,
\begin{equation}
E(m) = a \, \epsilon(m) + E_\mathrm{Extrapol.},
\end{equation}
in our extrapolations.
For all extrapolations, we replace $m$ in the notation introduced above by a list of those $m$ values for which energies are obtained for the extrapolation.

Entropy based entanglement measures defined for individual orbitals were introduced by Legeza and S\'olyom in 2003.\cite{legeza2003}
The single-orbital von Neumann entropy $s_i(1)$ for the $i$-th orbital can be calculated from the eigenvalues $w_{\alpha, i}$ of the one-orbital reduced density matrix,\cite{legeza2003,legeza2006,rissler2006}
\begin{equation}
 s_i(1) = - \sum_{\alpha = 1}^4 w_{\alpha, i} \ln w_{\alpha, i} ,
\end{equation}
where $\alpha$ labels all possible occupations (four in case of a spatial orbital).
The two-orbital entropy $s_{ij}(2)$ is obtained from the 16 eigenvalues $w_{\alpha, ij}$ of the two-orbital reduced density matrix,
\begin{equation}
 s_{ij}(2) = - \sum_{\alpha=1}^{16} w_{\alpha, ij} \ln w_{\alpha, ij} ,
\end{equation}
from which the \textit{mutual information} $I_{ij}$,\cite{rissler2006} which is a measure for the degree of entanglement between two orbitals, is then calculated as:
\begin{equation}
 I_{ij} = \frac{1}{2} [s_i(1) + s_j(1)-s_{ij}(2)](1-\delta_{ij}) .
\label{mutinf}
\end{equation}
Note that the prefactor and signs in Eq. (\ref{mutinf}) are given as implemented in \textsc{QCMaquis}.
The entanglement measures will be essential for the selection protocol of the active orbital space.
Our automated active orbital space selection is then realized with \textsc{Python} scripts based on selection criteria to be developed in the next section.

\section{Results and Discussion}

\subsection{Convergence of entanglement measures}

\begin{figure}[H]
 \includegraphics[width=\textwidth]{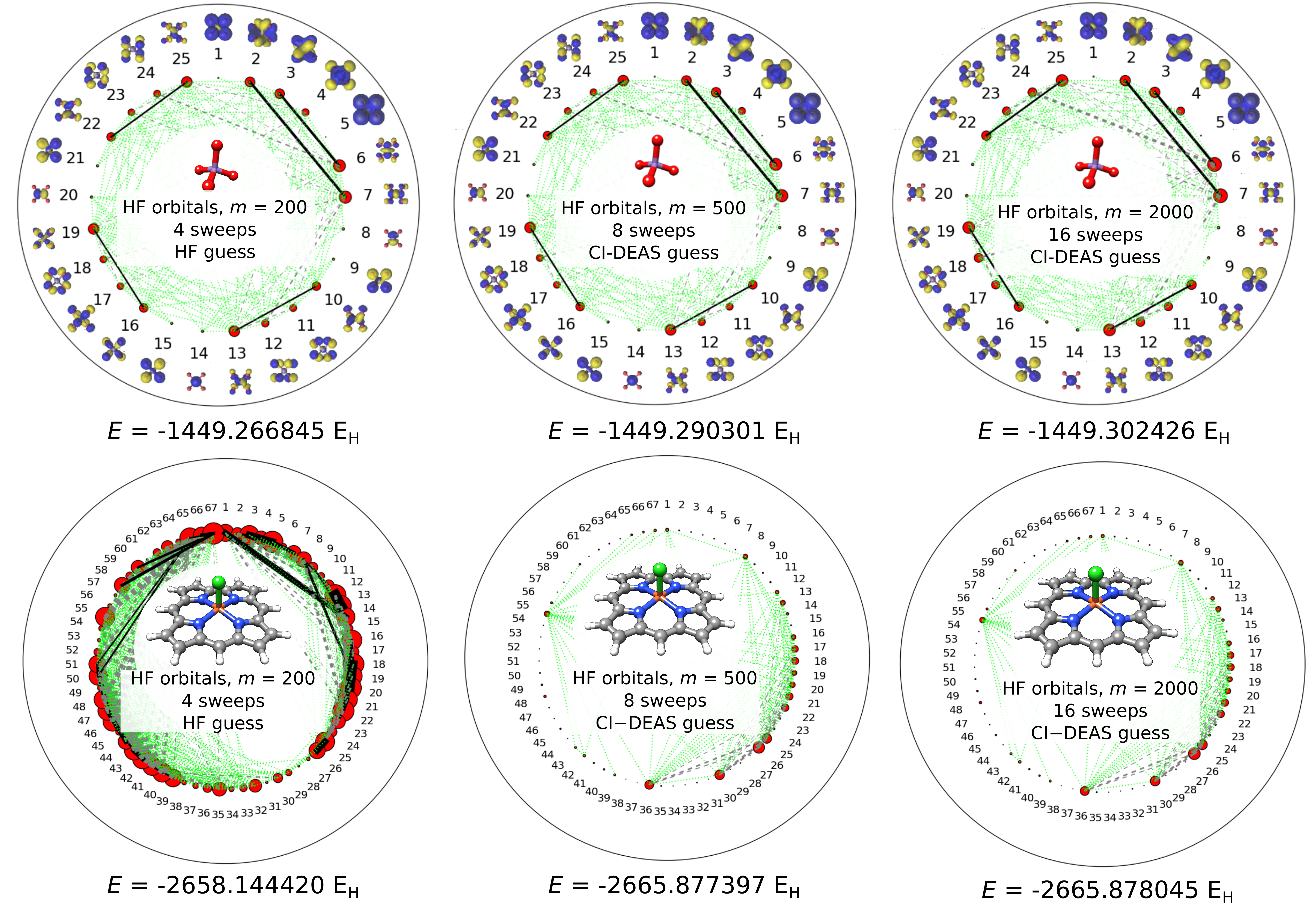}
 \caption{\small Entanglement diagrams calculated from DMRG wave functions obtained with different convergence protocols for MnO$_4^{-}$ and chloroiron corrole with Hartree--Fock orbitals. All orbitals are numbered and arranged on a circle. The area of the red circles assigned to each numbered orbital is proportional to the single-orbital entropy of the respective orbital. The line connecting two orbitals denotes their mutual information value. The lines in black indicate a value of $I_{ij}$ of at least 0.1, whereas dashed gray and green lines represent mutual information values of at least 0.01 and 0.001, respectively.  For the small CAS in the top panel it is feasible to show the orbitals, whereas this is not possible for the large active space in the lower panel. The molecular structures and DMRG settings are shown in the centers of the diagrams. Energies are given in Hartree, E$_\mathrm{H}$.}
 \label{ent_conv}
\end{figure}

The entropy based entanglement measures should be available at low computational cost in order to be useful.
Therefore, an approximate and hence fast calculation for a large active space of orbitals chosen around the Fermi level is performed to identify highly entangled orbitals that can then be selected for the final calculation.
We first investigate whether the entanglement measures calculated from an approximate wave function obtained in a partially converged, but fast DMRG calculation are qualitatively correct.
The upper panel of Figure \ref{ent_conv} shows entanglement diagrams containing a graphical representation of single-orbital entropies and mutual information for the permanganate ion with an active orbital space that includes all valence orbitals.
Three different DMRG settings are specified in the centers of these circular diagrams.
In the lower panel, a similar set of diagrams is shown for chloroiron corrole with a significantly larger CAS, i.e. a CAS(50,67).\\
The DMRG algorithm requires an initial guess of the environment states in the first iterations until one sweep is completed.\cite{moritz2006}
The importance of a suitable initial guess, such as the CI dynamically extended active space (CI-DEAS) approach\cite{legeza2004,legeza2010,molcaspaper}, is obvious from the results for chloroiron corrole for which a guess that only ensures the incorporation of the HF determinant fails to produce the correct entanglement pattern (see lower left corner of Figure \ref{ent_conv}).
Combined with a moderate number of renormalized block states of $m = 500$, the entanglement measures obtained with the CI-DEAS guess are hardly distinguishable from the converged results.
This observation equally holds for the single-orbital entropy and the mutual information.

For moderate active space sizes (see the results for MnO$_4^-$) even faster calculations with $m = 200$  and a HF initial guess result in sufficiently accurate entanglement measures despite the fact that the energy is still far from being converged.
We observe this independently of the type of orbitals and the degree of entanglement present in the molecule.
It was noted quite early that the convergence of DMRG calculations with HF orbitals tends to be slow\cite{chan2002} and we therefore present data for this most critical type of orbitals in Figure \ref{ent_conv}.
We observed the same results for all four types of orbitals independent on the number of orbitals in the active space (data for these other orbitals are not shown).
We recommend the CI-DEAS guess and $m = 500$ for active spaces with more than 30 orbitals to guarantee sufficiently accurate entanglement measures at low computational cost.
The calculation with the CAS(50,67) for chloroiron corrole employing the CI-DEAS guess and $m = 500$ with 8 sweeps requires a couple of hours of computational time on 16 cores of an Intel Xeon E5-2670 central processing unit and gives reliable entanglement information.
This entanglement information then allows for an automated active orbital space selection shown in the following sections so that the benefit more than outweighs the additional computational cost.
In general, the entanglement measures from partially converged DMRG calculations closely resemble the converged results so that an active space selection based on these approximate calculations is justified.

\subsection{Orbital selection protocols}

We now define criteria based on entanglement information for the automated selection of compact active orbital spaces.
Highly entangled orbitals are essential to construct a reliable zeroth-order wave function in a CAS approach and must be included in the active orbital space.\cite{reiher2012}
The single-orbital entropy and the maximum value of the mutual information serve as measures for the degree of entanglement of individual orbitals.\\
Global selection thresholds can be defined, for example, as a fraction of the theoretical maximum values (e.g., $s(1)_\mathrm{max.} = \ln(4) \approx 1.4$ in a spatial-orbital basis with four possible occupations).
To account for the varying degree of orbital entanglement in different molecules, however, we choose our selection criteria with respect to the 
maximum value of the single-orbital entropy and mutual information obtained from the same calculation rather than with respect to the theoretical maximum.
This ensures transferability of the threshold as the degree of entanglement strongly varies with the molecule and the orbital basis under consideration.
A global selection threshold defined as a fraction of the theoretical maximum can still be advantageous in a first step to decide whether a molecule requires a multi-configuration ansatz or whether a single-configuration method is applicable.
The electronic structure of all molecules considered here was found to be of multi-configurational character in previous works (cited below).
We analyzed the maximum single-orbital entropies for all molecules and orbital bases in order to find a threshold that identifies all these cases as multi-configurational.
From this, we conclude that a conservative global threshold of $0.1\cdot s_i(1)_\mathrm{max.} \approx 0.14$ for at least one orbital indicates the multi-configurational character of a wave function and includes even borderline cases like the CrF$_6^{3-}$ complex discussed  below.

For the final active space, we then select all orbitals whose single-orbital entropy is higher than a fraction of the maximum value found for one of the $s_i(1)$ in the calculation under consideration.
This fraction is dynamically determined for each calculation as described below.
A threshold based on the maximum element of the mutual information matrix is also investigated.
We choose the permanganate ion to illustrate the approach, because it is a difficult molecule for electronic structure calculations despite its small size.
A suitable active space for MnO$_4^-$ was under debate for some time and the general rules for the active orbital space selection mentioned in the Introduction had to be extended for this covalently bonded transition-metal complex.\cite{pierloot2001,veryazov2011}
All four different types of starting orbitals --- HF, split-localized, CASSCF orbitals from a small CAS(10,10)-SCF calculation and partially converged DMRG-SCF orbitals from the whole valence space --- are considered.

\begin{figure}[H]
 \includegraphics[width=\textwidth]{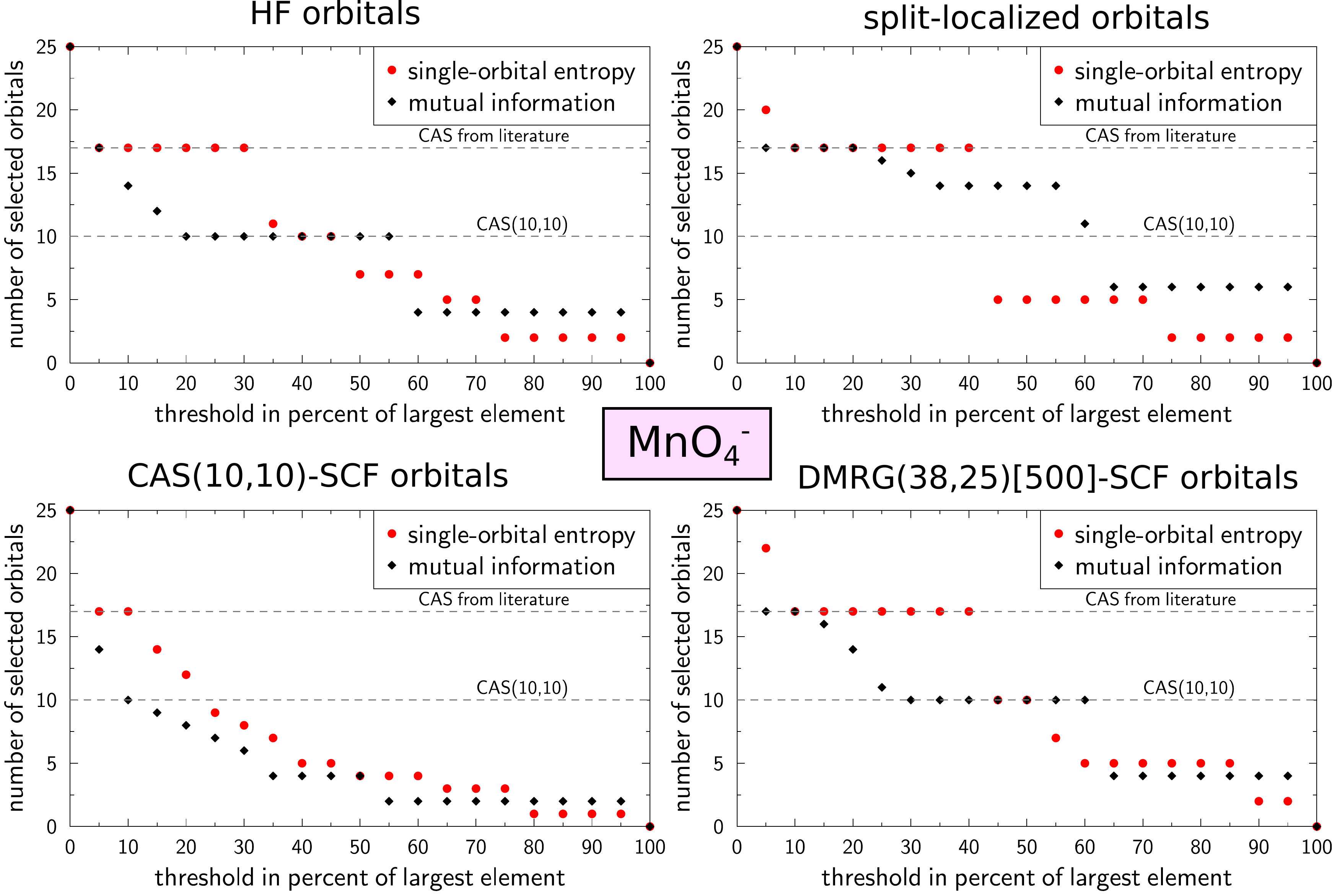}
 \caption{\small Dependence of the number of selected orbitals on the threshold values for the single-orbital entropy (red) and the mutual information (black), shown for HF, split-localized, CAS(10,10)-SCF and DMRG(38,25)[$500$]-SCF orbitals of the MnO$_4^-$ ion. On the $x$-axis, the orbital selection threshold is given as a fraction of the maximal value for each calculation.}
\label{thresh_mno4}
\end{figure}

For the identification of a reliable threshold for the orbital selection we apply threshold diagrams as shown in Figure \ref{thresh_mno4}.
On the $x$-axis the orbital discarding threshold is varied from 0 to 100~\% of the largest element of either the single-orbital entropy or the mutual information matrix, while the number of selected orbitals according to that threshold is displayed on the $y$-axis.
A threshold of 0~\% then leads to a selection of all orbitals (here 25 orbitals on the left of each diagram in Figure \ref{thresh_mno4}), whereas a threshold of 100~\% of the maximum element (on the right of each diagram in Figure \ref{thresh_mno4}) selects no orbital for the final converged calculation.
The threshold diagrams contain in general one or more plateaus (defined as a slope of zero over a threshold range of at least 10~\%) that identify subsets of orbitals with similar degree of entanglement.
In the threshold diagram for HF orbitals (upper left panel of Figure \ref{thresh_mno4}) for example, a subset of 17 orbitals has single-orbital entropies of at least 30~\% of the maximum $s_i(1)$ value (right end of the first plateau) and only when the threshold is lowered to less than 5~\% (left end of the first plateau) more orbitals (all 25) are selected.

In the same threshold diagram, the mutual information threshold leads to a different plateau with only ten orbitals.
The set of 17 orbitals is identical to the active orbital space recommended in the literature, whereas a CAS with only ten orbitals was considered insufficient.\cite{pierloot2001,veryazov2011} 
Since the mutual information threshold in general leads to plateaus with a lower --- and in this case insufficient --- number of orbitals, we conclude that a threshold based on the single-orbital entropy is more reliable and will focus on this in the following.

The correct CAS  evolves here naturally from the discarding threshold that is dynamically determined from the plateaus in the single-orbital entropy threshold diagrams for all four types of orbitals.
The definition of this threshold is, however, less clear for the CAS(10,10)-SCF orbitals (lower left diagram of Figure \ref{thresh_mno4}), where no plateaus can be identified.
The small CAS chosen for the generation of these orbitals includes ten orbitals around the Fermi level.
In this case, it is possible to exclude eight orbitals with very small single-orbital entropy, leaving the same 17 orbitals in the final active space as for the other types of orbitals.
Interestingly, the ten orbitals of the initial CAS do not give rise to a plateau in the threshold diagram and we may conclude that orbitals from a small CAS calculation do not introduce a bias towards the initial active orbitals.
This indicates that the choice of orbitals to be included in the active space is arbitrary to some degree, because even orbitals from a CAS known to be insufficient lead to the selection of the correct set of active orbitals in an automated approach.

From the threshold diagrams in Figure \ref{thresh_mno4} and those in the following sections we can identify two different entanglement patterns:
Whenever plateaus occur in the diagrams, highly entangled orbitals that have to be \textit{included} in the final active space can be identified, whereas weakly entangled orbitals will be \textit{excluded} if no such plateaus are present.

\begin{figure}[H]
 \includegraphics[width=\textwidth]{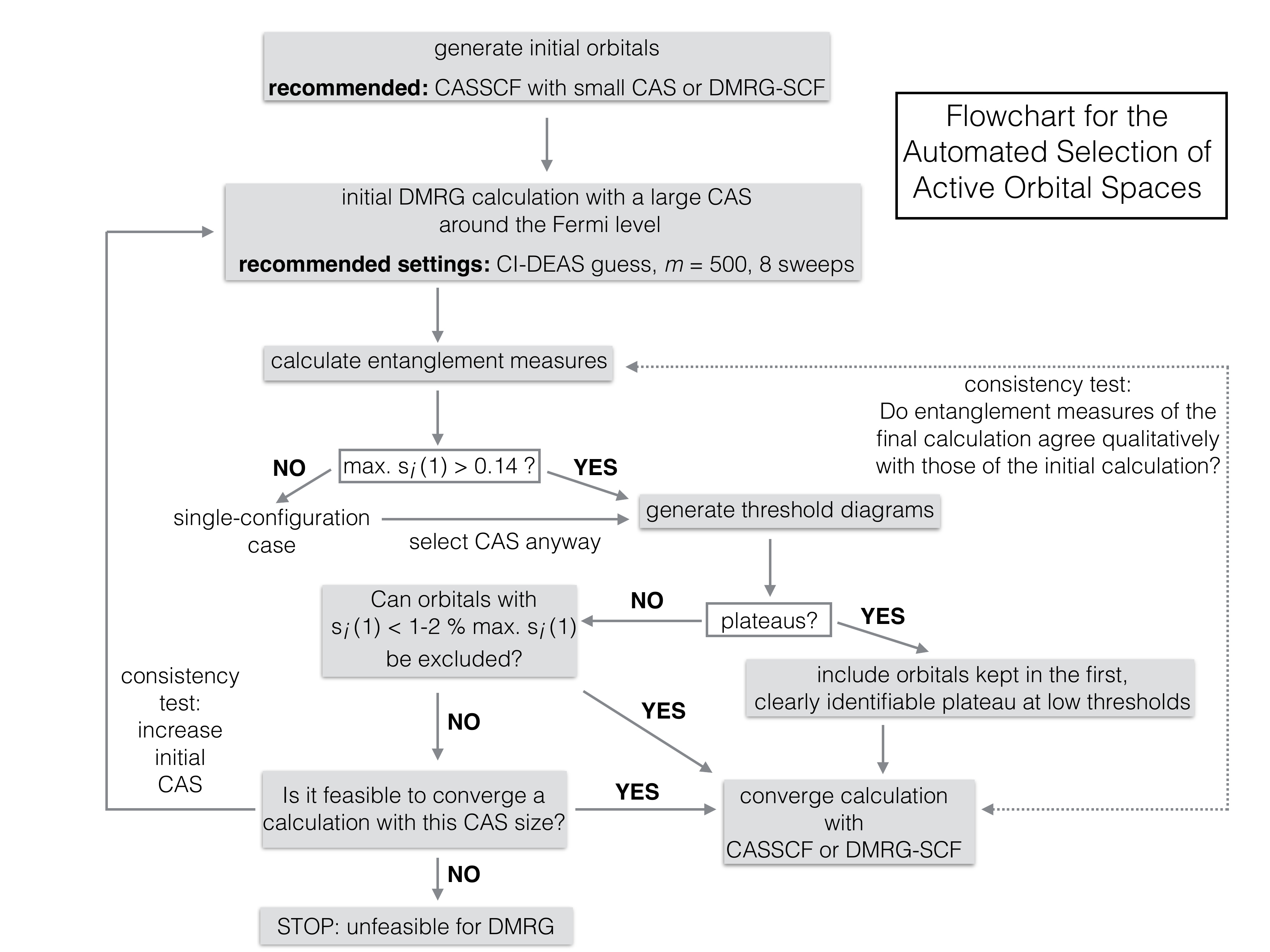}
 \caption{\small The flowchart illustrates the procedure of the CAS selection.}
\label{flowchart}
\end{figure}

Based on these results, we propose the following protocol for orbital selection from the initial, exploratory large-CAS DMRG
calculation that is now solely based on the single-orbital entropy depicted in Figure \ref{flowchart}:
\begin{enumerate}
\item We will assign a multi-configuration character of the wave function if at least one single-orbital entropy is higher than 0.14.
\item If we observe a plateau structure in the threshold diagrams, the corresponding highly entangled orbitals will be identified and included in the final active space.
\item If no plateau structure is observed (as is the case for the CAS(10,10)-SCF orbitals in MnO$_4^-$), we will exclude orbitals whose single-orbital entropy is lower than 1-2~\% of the maximum single-orbital entropy.
\item If all orbitals are to be selected, an even larger CAS is selected for the initial DMRG calculation to probe whether additional orbitals are required for the final CAS.
\item A comparison of the entanglement information extracted from the converged selected-CAS calculation with that of the initial calculation allows us to assess the consistency of the selection procedure.
\end{enumerate}

Apart from the initial orbital generation all steps are automated.
A selection based on a fixed threshold may also be applied.
From the results above, a threshold of 10~\% of the maximum $s(1)$-value would lead to the same active orbital selection as the procedure introduced above (for all types of orbitals).

We emphasize that we define our selection protocol based on a comparison to active spaces recommended in the literature.
That this protocol yields reliable results for, e.g., relative energies will be demonstrated in Sections \ref{salen} and \ref{copper}.
Figure \ref{thresh_mno4} further suggests that all types of orbitals studied here are equally suited for the entanglement based selection of active orbital spaces.
In the following, however, we do not further consider HF and split-localized orbitals.
The delocalized nature of the former makes a preselection of orbitals hard for large molecules and can also lead to slow DMRG convergence in the final calculation.
The latter type of orbitals overcomes this issue but at the expense of loosing point-group symmetry so that calculations are computationally more demanding.
Moreover, the localization procedure is not always straightforward, especially for the virtual orbitals.
Hence, we focus on CASSCF orbitals from a minimal CAS and on partially converged DMRG-SCF orbitals in the following.
The latter proved to be valuable in cases where the initial CAS did not exceed 40 orbitals as shown in the following sections.
Note that the generation of these orbitals took less than a day (on 16 cores of an Intel Xeon E5-2670 central processing unit) in the worst case of Cu$_2$O$_2^{2+}$ with 36 orbitals.

\subsection{Covalent and non-covalent interactions}
\label{chrome}

The degree of covalency in the metal-ligand bonds heavily affects the number of orbitals to be included in the active space. 
This was demonstrated in the case of the chromium hexafluoro complex, where a change of the total charge changes the degree of covalency of the coordination bonds.\cite{pierloot2001}
While the bonds in the trianionic species CrF$_6^{3-}$ are almost purely ionic, the neutral species CrF$_6$ is highly covalent.
This is reflected in the natural orbital occupation numbers (NOONs) of restricted active-space SCF (RASSCF) orbitals,\cite{pierloot2001} where the NOONs of all 23 valence orbitals in the neutral complex differ significantly from 2 and 0 for occupied and unoccupied orbitals, respectively.

Based on these RASSCF occupation numbers, the authors of Ref. \citenum{pierloot2001}  conclude that a suitable choice for the active space of the neutral complex comprises all 23 valence orbitals, while a CAS(13,10) including the whole set of $t_{2_g}$, $e_g$, and their anti-bonding counterparts is a valid choice for the anionic complex.
The latter suggestion, however, is hard to confirm based on the occupation numbers alone, because all orbitals are either fully occupied or unoccupied, as can be seen from the RASSCF data of Table \ref{occnum_crf6}.
To investigate whether our entanglement based selection criteria reflect the degree of covalency, we performed DMRG($N$,23)[$1000$]-SCF calculations for both the trianionic ($N$ = 39) and the neutral ($N = 36$) species.

\begin{figure}[H]
 \includegraphics[width=\textwidth]{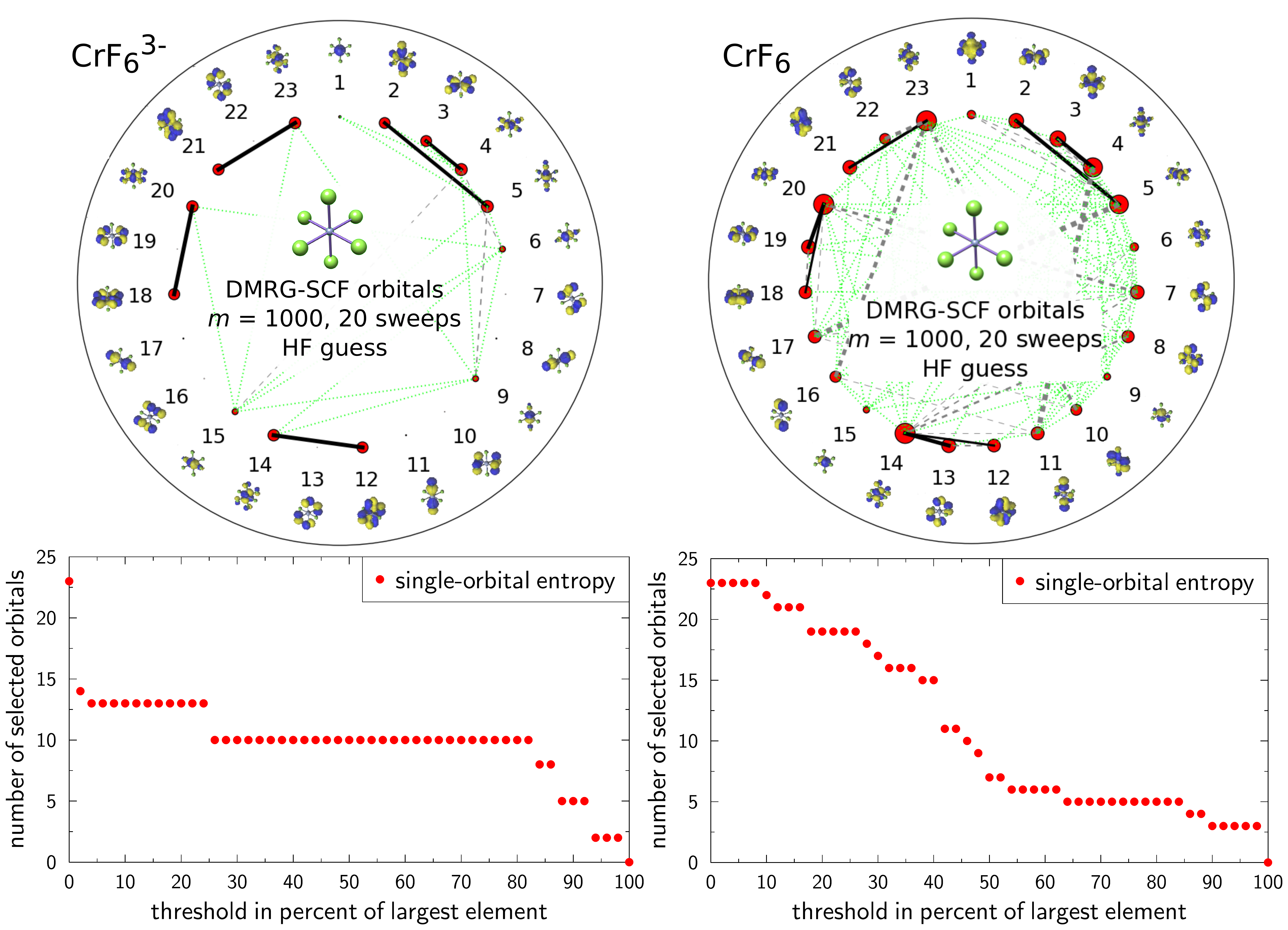}
 \caption{\small Upper panel (notation as in Figure \ref{ent_conv}): entanglement diagrams for all 23 DMRG($N$,23)[$1000$]-SCF valence orbitals of CrF$_6^{3-}$ ($N = 39$) and CrF$_6$ ($N = 36$). Lower panel: Single-orbital entropy threshold diagrams for the determination of the number of orbitals to be included in the active space.}
 \label{crf6_fig}
\end{figure}

As can be seen from Figure \ref{crf6_fig}, the orbital entanglement is significantly different for both species.
While in the anionic complex only the set of $t_{2_g}$, $e_g$, $t_{2_g}^*$, and $e_g^*$ orbitals shows a significant degree of entanglement, all orbitals are highly entangled in the neutral species.

The completely different degree of entanglement is clearly reflected in the single-orbital entropy threshold diagrams in the lower panel of Figure \ref{crf6_fig}.
For the anionic complex, the number of orbitals with a high single-orbital entropy drops at very low fractions of the largest value and is then constant over a large range.
The threshold diagram in the lower left corner of Figure \ref{crf6_fig} shows two extended plateaus that would allow one to choose
two different CAS. We therefore identify a set of ten orbitals with up to 80 \% of the highest single-orbital entropy and may select either these 
orbitals (as previously proposed) 
or a slightly larger set of 13 orbitals considering the first plateau at low thresholds.
In view of the (second) plateau at higher fractions of the maximum $s_i(1)$ value extending over a large threshold 
range, the set of ten orbitals is chosen to yield a compact CAS.
Note, however, that plateaus at very high thresholds would, in general, select too few orbitals and can therefore be safely discarded.
From the molecules investigated in this study we deduce that plateaus in threshold diagrams are only meaningful up to a threshold of 60~\%.

The entanglement information pattern is totally different for the neutral complex, where no such plateau is observed in the threshold diagrams.
According to the selection protocol, we attempt to exclude all orbitals with a single-orbital entropy of less than 1-2~\% of the highest value in this case.
Since this does not apply to any of the orbitals, we consequently have to select all 23 valence orbitals, exactly as proposed in Ref. \citenum{pierloot2001}.

A comparison with selection criteria based on NOONs can be made in view of the data in Table \ref{occnum_crf6}.
The DMRG-SCF NOONs suggest to choose the same set of orbitals as the entanglement based selection for both complexes, while the RASSCF NOONs would erroneously make the anionic species a case suitable for single-reference methods.
Therefore, DMRG-SCF NOONs will lead to the same conclusions as those drawn from the entanglement information for both complexes with similar constitution but very different electronic structure.

\begin{table}[H]
\caption{\small NOONs for the anionic and neutral CrF$_6$ complexes obtained from DMRG[$1000$]-SCF calculations and RASSCF results taken from Ref. \citenum{pierloot2001}.}
\begin{center}
\begin{tabular}{lcccc}
\hline
\hline
&\multicolumn{2}{c}{CrF$_6^{3-}$ / CAS(39,23)}&\multicolumn{2}{c}{CrF$_6$ / CAS(36,23)}\\
orbital & RASSCF & DMRG-SCF & RASSCF$^a$ & DMRG-SCF \\
\hline
$t_{2_u}$    & 6.00 & 5.96 & 5.89 & 5.91 \\
$t_{1_g}$    & 6.00 & 6.00 & 5.91 & 5.90 \\
$a_{1_g}$   & 2.00 & 2.00 & 1.98 & 1.97 \\
$t_{1_u}$    & 6.00 & 6.00 & 5.91 & 5.89\\
$t_{1_u}$    & 6.00 & 6.00 & 5.85 & 5.83 \\
$t_{2_g}$    & 6.00 & 5.83 & 5.84 & 5.78 \\
$e_{g}$       & 3.99 & 3.91 & 3.88 & 3.76 \\
$t_{2_g}^*$ & 3.00 & 3.18 & 0.46 & 0.53 \\
$e_{g}^*$    & 0.01 & 0.11 & 0.28 & 0.39 \\
\hline
\hline
\end{tabular}
\end{center}
\label{occnum_crf6}
\end{table}

\subsection{Spin-state energetics of oxo-Mn(salen)}
\label{salen}

The automated orbital selection should also select an active space that consistently describes several electronic states such as spin states of a molecule.
We choose oxo-Mn(salen) as an example to show that this is accomplished.
This complex catalyzes the enantioselective epoxidation of unfunctionalized olefins\cite{jacobsen1990,katsuki1990} and was intensively studied with \textit{ab initio} methods.\cite{ivanic2004,sherill2006,gagliardi2011,wouters2014,chan2015}
The CASSCF studies of Ivanic \textit{et al.}\cite{ivanic2004} and Sears \textit{et al.}\cite{sherill2006} applied a very limited size of the active space, whereas the later studies by Ma \textit{et al.}\cite{gagliardi2011},  Wouters \textit{et al.}\cite{wouters2014} and Olivares-Amaya \textit{et al.}\cite{chan2015} were able to overcome that limitation through the generalized active space method in the former study and DMRG in the latter two studies.
The DMRG study of Wouters \textit{et al.} is relevant to the context of our work, because the authors proposed a recipe to select an active space based on the occupation numbers of an initial (expensive) DMRG[$3000$]\#HF calculation for the singlet state. 
The active space contained a large number of 40 to 45 orbitals (depending on the basis set) around the Fermi level.
The final selection led to a CAS(28,22) and these orbitals were further optimized for each spin state separately.
So far, this occupation number based approach proposed by Wouters \textit{et al.} has only been tested for this particular case.

\begin{figure}[H]
 \includegraphics[width=\textwidth]{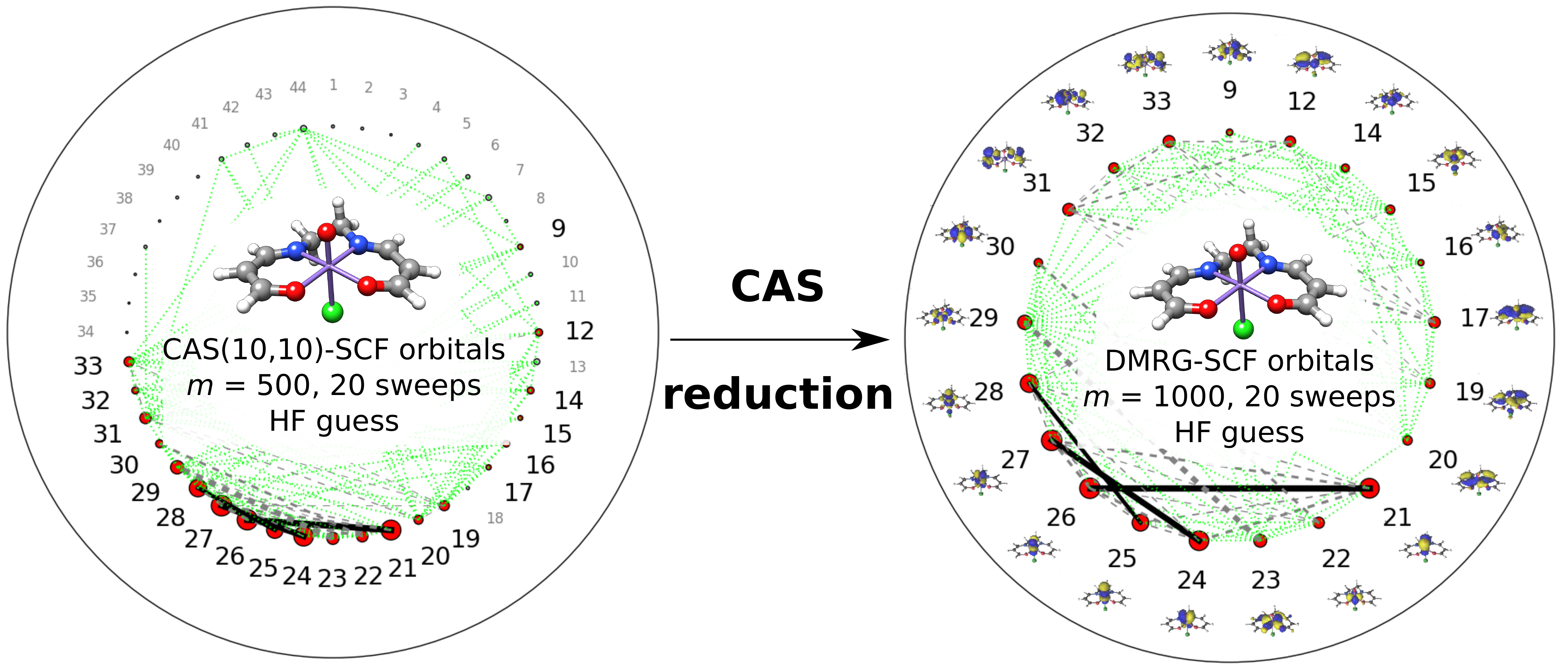}
 \caption{\small Entanglement diagrams for the singlet state of oxo-Mn(salen). The left diagram shows the entanglement information of the 44 orbitals around the Fermi level. On the right, the entanglement information of a converged calculation with the selected final subset of these orbitals is shown (notation as in Figure \ref{ent_conv}).}
 \label{salen_sel}
\end{figure}

In this work, we start from CAS(10,10)-SCF orbitals, where the active space consists of all orbitals with significant metal $3d$-character.
DMRG(50,44)[$500$]\#CAS(10,10)-SCF calculations for each spin state (singlet, triplet, and quintet) were then performed to obtain the entanglement measures.
Since the entanglement of the orbital basis can be different for each spin state and a common CAS for all spin states is preferable in order to avoid artifacts, we construct our CAS from the union of orbitals selected for each spin state.
All orbitals with a single-orbital entropy of at least 10 \% of the highest value of each spin state are selected, leading to a final unified CAS(26,21).
These orbitals were then optimized in a DMRG(26,21)[$1000$]-SCF calculation for each spin state separately.
We emphasize that in the study by Wouters \textit{et al.}\cite{wouters2014}, the selection was solely based on a calculation for the singlet state so that the orbital selection might be biased.
In our case, that restriction would have led to an even smaller CAS(20,18).
In Figure \ref{salen_sel}, the selection of the 21 orbitals is shown exemplarily for the calculation of the singlet state. 
Although orbitals 9, 15, and 17 would not have been selected by the 10 \% criterion for the singlet state, they were selected on the basis of the entanglement information of either the triplet or quintet wave function and were therefore included in the unified CAS. 

\begin{figure}[H]
\begin{center}
\includegraphics[width=0.8\textwidth]{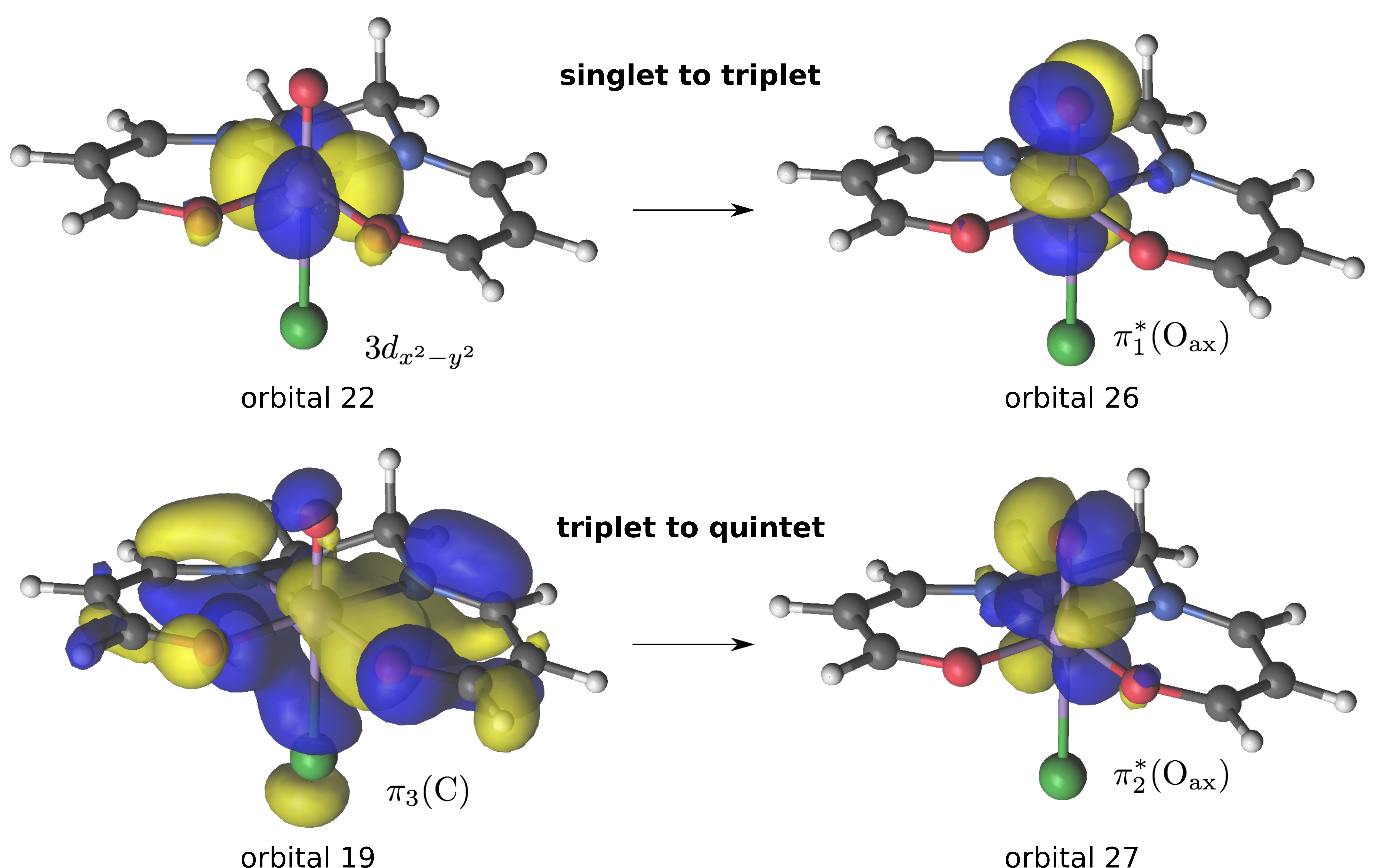}
\end{center}
\caption{\small DMRG(26,21)[$1000$]-SCF orbitals involved in electronic transitions leading to the different spin states. The singlet-triplet transition can be interpreted as an  $3d_{x^2-y^2}$ $\rightarrow $ $\pi_1^*$(O$_\mathrm{ax}$) excitation, while the triplet-quintet transition can be described as a  $\pi_3$(C) $\rightarrow$ $\pi_2^*$(O$_\mathrm{ax}$) excitation.}
\label{orb_salen}
\end{figure}

\begin{figure}[H]
\begin{center}
\includegraphics[width=0.8\textwidth]{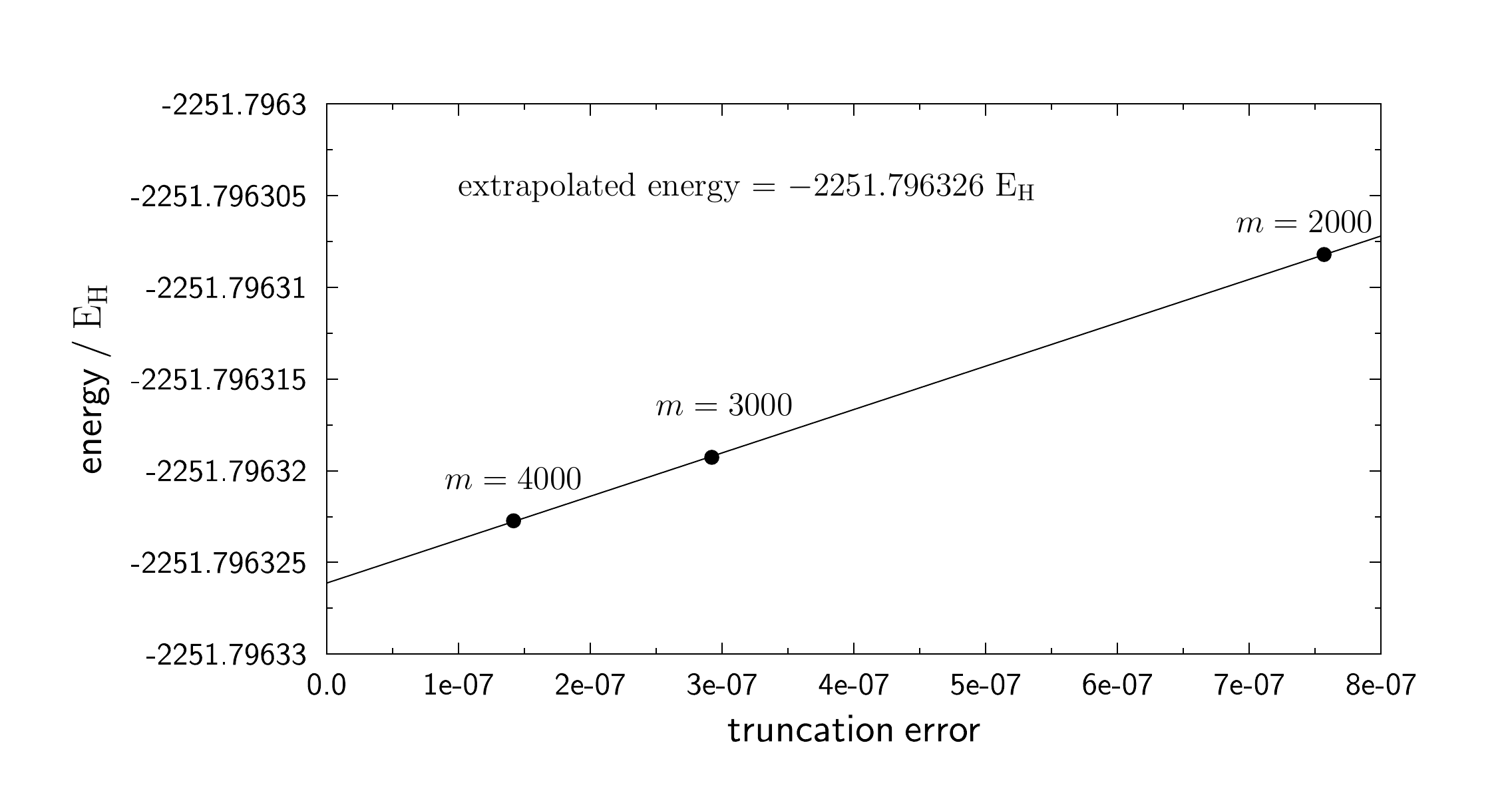}
\end{center}
\caption{\small Extrapolation of the DMRG energies (in Hartree, E$_\mathrm{H}$) with respect to the truncation error $\epsilon$ in Eq. (\ref{truncerr}) for 
$m = 2000, 3000$, and 4000 for the singlet state of oxo-Mn(salen).}
\label{extrapol}
\end{figure}

From the NOONs of the final DMRG calculations driven to full convergence for all three spin states (collected in Table \ref{occnum_salen}), the excitations that lead to the different spin states can be interpreted in a simple orbital picture.
The generation of the triplet state can then be understood as the promotion of one electron from the $3d_{x^2-y^2}$-orbital to the anti-bonding $\pi_1^*$(O$_{ax}$)-orbital.
The quintet state can be generated by a subsequent excitation of an electron from the $\pi_3$(C)-orbital (located mainly on the salen ligand) to the second anti-bonding $\pi_2^*$(O$_{ax}$)-orbital as shown schematically in Figure \ref{orb_salen}.
This observation is in accord with the findings of the DMRG study by Wouters \textit{et al.}\cite{wouters2014} but contrasts earlier studies\cite{ivanic2004,sherill2006}, where the quintet is interpreted as an excitation of an electron from a bonding $\pi$(O$_{ax}$)-orbital into its anti-bonding counterpart.
The different electronic structure of the quintet state in the earlier studies\cite{ivanic2004,sherill2006} can be explained by the fact that all $\pi$-orbitals on the salen ligand were artificially kept doubly occupied because of restrictions of the active space size in traditional CASSCF. 
Consequently, the quintet state of these studies with a severely limited CAS size is 40-45~kcal/mol higher in energy than the singlet,\cite{ivanic2004,sherill2006} while in our calculations this energy difference amounts to only 25.6~kcal/mol.
These final relative energies were obtained by extrapolating DMRG(26,21)\-[$2000,3000,4000$]\-\#DMRG(26,21)[$1000$]-SCF energies to a truncation error of zero for each spin state (see Figure \ref{extrapol}).
A comparison between the energies obtained in this work and the previous DMRG study is given in Table \ref{spin-state-comp}.
In contrast to Wouters \textit{et al.}\cite{wouters2014} who found a triplet ground state that is $-5.3$~kcal/mol lower in energy than the singlet state, we find the latter to be about 0.5 kcal/mol lower than the triplet so that the two states are quasi-degenerate.
Our result, however, agrees well with the DMRG study of Ref. \citenum{chan2015} that also found a singlet ground state with a singlet-triplet splitting of 0.4 kcal/mol from a manually selected CAS(32,24).

As dynamical correlation effects were not taken into account in either study, the correct energetic order cannot be determined.
Our quintet state, however, lies almost twice as high in energy as in the previous calculations (24.6 kcal/mol compared to 12.1 kcal/mol) and is now in between the previous DMRG result and the earlier CASSCF results (40-45 kcal/mol).

We further note that our choice of the active orbital space, although including one orbital less than Wouters \textit{et al.}\cite{wouters2014}, gives lower absolute energies.
The consistently lower total electronic energy of about 30~kcal/mol for all three spin states in our calculations compared to Ref. \citenum{wouters2014} cannot be explained by a bias from selecting active orbitals based on a calculation for the singlet state only.
Since we used the same structure and basis set, our active space appears to capture more correlation energy than the larger CAS described in Ref. \citenum{wouters2014}.

\begin{table}[H]
\caption{\small Natural orbital occupation numbers for the selected orbitals of oxo-Mn(salen) after DMRG(26,21)[1000]-SCF orbital optimization for each spin-state. Values in bold correspond to the singlet-triplet [$3d_{x^2-y^2}$ $\rightarrow $ $\pi_1^*$(O$_\mathrm{ax}$) ] and triplet-quintet [$\pi_3$(C) $\rightarrow$ $\pi_2^*$(O$_\mathrm{ax}$) ] transitions as depicted in Figure \ref{orb_salen}.}
\begin{center}
\begin{tabular}{llllllllll}
\hline
\hline
Nr. & type & $^1 A$ & $^3 A$  & $^5 A$ & Nr. & type & $^1 A$ & $^3 A$  & $^5 A$ \\
\hline
9   & $3p$(Mn)             &1.99 &1.99 &1.99 & 24 & $\pi_2$(O$_\mathrm{ax}$)      &1.86 &1.78 &1.94 \\
12 & $\pi_1$(C)            &1.96&1.96 &1.95 & 25 &$\sigma$(O$_\mathrm{ax}$)    &1.92 &1.91 & 1.90\\
14 & $3p$(Mn)              &1.99&1.99 &1.99 & 26 & $\pi_1^*$(O$_\mathrm{ax}$)   &\textbf{0.16} &\textbf{1.04} &\textbf{1.04} \\
15 & $\sigma$(Mn-N)  &1.99&1.98 &1.99 & 27 & $\pi_2^*$(O$_\mathrm{ax}$)   &\textbf{0.15} &\textbf{0.23} & \textbf{1.04}\\
16 &$3p$(Mn)              &2.00 &1.99 &2.00 & 28 & $\sigma^*$(O$_\mathrm{ax}$)&0.10 &0.11 & 0.11\\
17 &$\pi_2$(C)             &1.98 &1.96 & 1.91& 29 & $\sigma^*$(salen)      &0.05 &0.08 & 0.04\\
19 & $\pi_3$(C)            &\textbf{2.00} &\textbf{2.00} &\textbf{1.00} & 30 & n.a.$^a$                     &0.02 &0.02 & 0.02\\
20 &$\pi_4$(C)             &1.96 &1.99 &2.00 & 31 & $\pi_1^*$(C)               &0.04 &0.05 & 0.09\\
21 &$\pi_1$(O$_\mathrm{ax}$) &1.86 &1.95 &1.94 & 32 &n.a.$^a$                      &0.01 &0.01 & 0.02\\
22 &$3d_{x^2-y^2}$      &\textbf{1.97} &\textbf{1.00} &\textbf{1.00} & 33 & $\pi_2^*$(C)              &0.04 &0.04 & 0.05\\
23 &$\sigma$(salen)     &1.96 &1.94 &1.98 & & & & & \\
\hline
\hline
\multicolumn{10}{l}{$^a$ The character of the orbital changed during the optimization and is}\\
\multicolumn{10}{l}{different for each spin state.}\\
\end{tabular}
\end{center}
\label{occnum_salen}
\end{table}

\begin{table}[H]
\caption{\small Total DMRG(26,21)[2000,3000,4000]\#DMRG(26,21)[1000]-SCF electronic energies (top three rows, in Hartree) and relative energies (two rows at the bottom, in kcal/mol) for the lowest singlet, triplet, and quintet states of oxo-Mn(salen). Values in parentheses are taken from Ref. \citenum{wouters2014}.}
\begin{center}
\begin{tabular}{lc}
\hline
\hline
 state & electronic energy \\
\hline
$^1 A$                 &  -2251.796326 (-2251.7509) \\
$^3 A$                 &  -2251.795396  (-2251.7593)\\
$^5 A$                 &  -2251.757044  (-2251.7316)\\
$^3 A$ -- $^1 A$   &  0.6  (-5.3)\\
$^5 A$ -- $^1 A$   &  24.6 (12.1) \\
\hline
\hline
\end{tabular}
\end{center}
\label{spin-state-comp}
\end{table}

\subsection{The inter-conversion of two isomers of Cu$_2$O$_2^{2+}$}
\label{copper}
In this section, we show that the automated active-orbital selection is also possible along a reaction coordinate.
A model system for which this can be investigated is the inter-conversion of the bis($\mu$-oxo) isomer of Cu$_2$O$_2^{2+}$ to its peroxo form.
This isomerization was intensively studied with a multitude of quantum-chemical methods\cite{cramer2006,gagliardi2008,legeza2011,neese2011,yanai2009,scuseria2012}. 
It proved to be a difficult case for traditional CASSCF because of the limitation of the CAS size.
This could not completely be overcome by RASSCF with perturbation theory of second order (RASPT2) calculations when compared to the best available \textit{ab initio} data provided by the coupled-cluster variant CR-CCSD(TQ)$_L$.\cite{gagliardi2008}

\begin{figure}[H]
 \includegraphics[width=\textwidth]{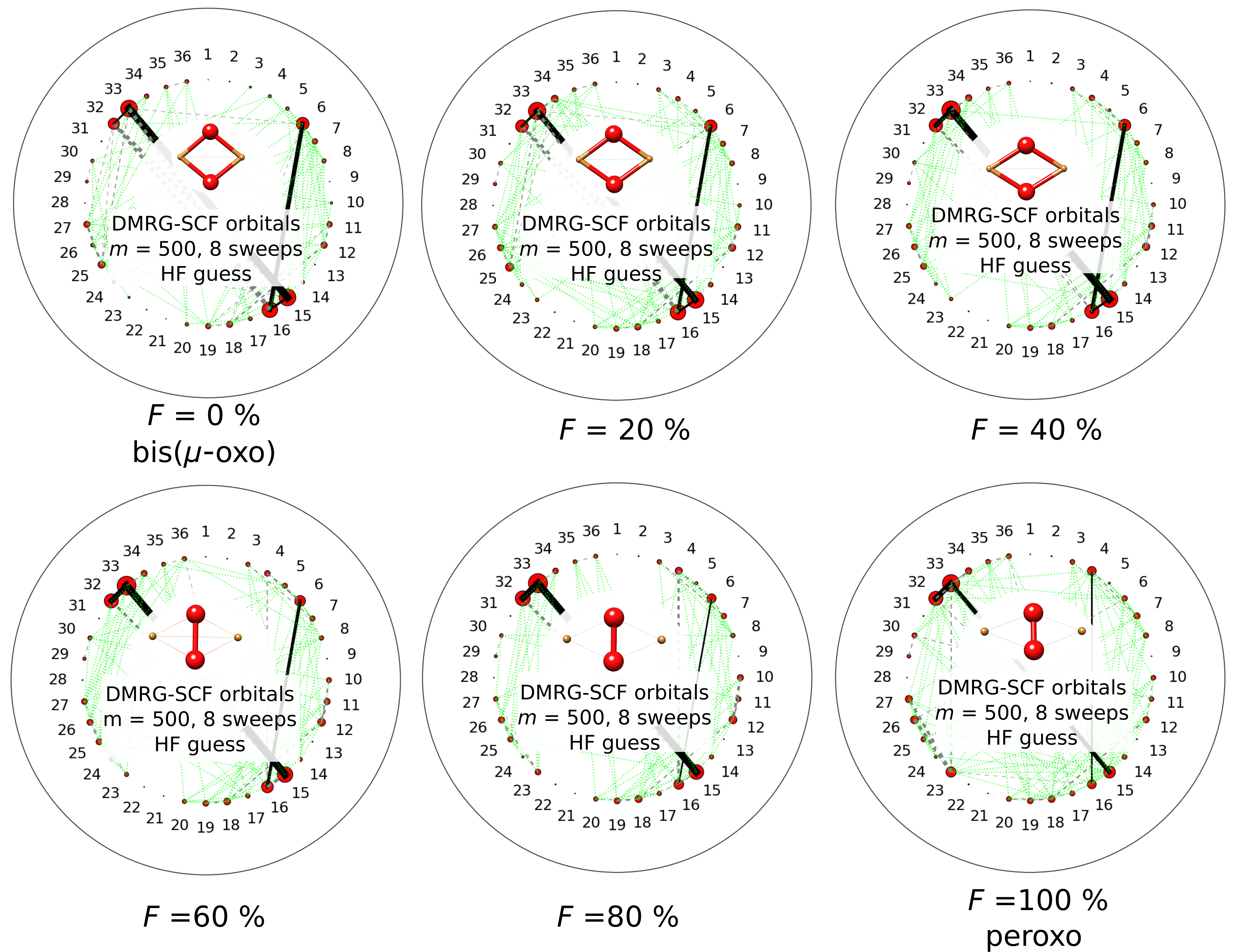}
 \caption{\small Entanglement diagrams for the partially converged DMRG(48,36)[$500$]-SCF calculations for six structures along the isomerization coordinate of Cu$_2$O$_2^{2+}$ (notation as in Figure \ref{ent_conv}).}
 \label{mutinf_cu}
\end{figure}

For the calculation of relative energies along the isomerization coordinate, DMRG(48,36)[$500$]-SCF calculations are performed that cover the whole valence space.
Six points are chosen as in Ref.~\citenum{cramer2006}, where the Cu-Cu and O-O distances are 2.8 and 2.3~\AA, respectively, for the bis($\mu$-oxo) structure in $D_{2h}$ symmetry.
In the peroxo structure, these distances are 3.6 and 1.4~\AA, respectively.
The distances for the structures along the isomerization coordinate are then calculated according to
\begin{equation}
q_i(F) = q_i(\mathrm{bis(}\mu\mathrm{-oxo)}) + \frac{F}{100}[q_i(\mathrm{peroxo})-q_i(\mathrm{bis(}\mu\mathrm{-oxo)})],
\end{equation}
where $q_i$ is either the O-O or the Cu-Cu bond length and $F$ is the fraction of progress along the coordinate.

\begin{figure}[H]
 \includegraphics[width=\textwidth]{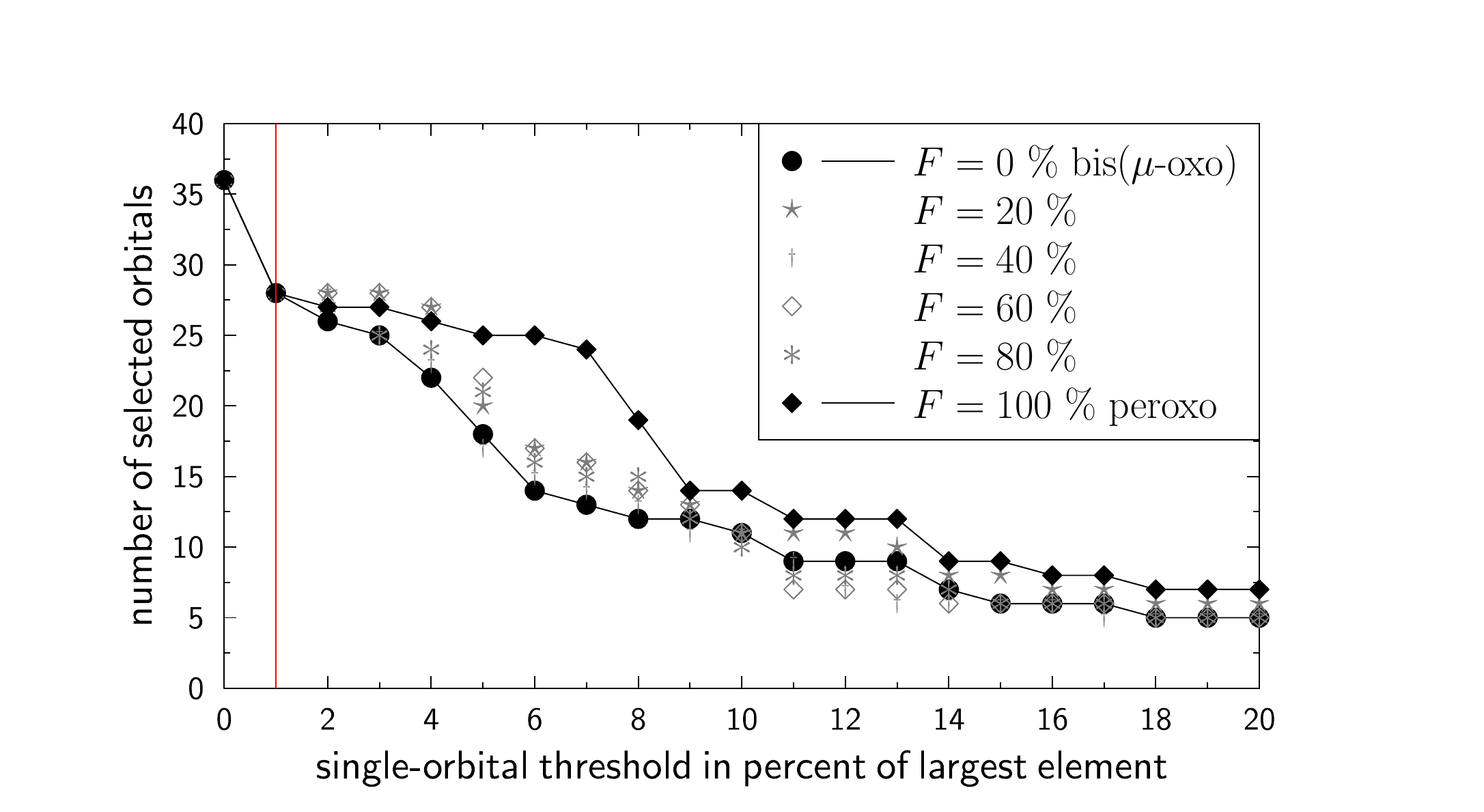}
 \caption{\small Number of selected orbitals from a varying single-orbital entropy threshold given in percent of the largest single-orbital entropy for six structures along the isomerization coordinate of Cu$_2$O$_2^{2+}$. The entanglement measures were extracted from DMRG(48,36)[500]-SCF calculations. The red line marks a threshold that selects a set of 28 orbitals for the final calculation.}
 \label{thresh_cu}
\end{figure}

The entanglement measures obtained are depicted in Figure \ref{mutinf_cu}.
Figure \ref{thresh_cu} collects the threshold diagrams of all structures obtained from the initial DMRG calculations.
When a discarding threshold of 0~\% (very left in Figure \ref{thresh_cu}) is employed, all 36 valence orbitals are selected for the active space. For all six structures, the number of selected orbitals is reduced to 28 at a threshold of 1~\% (red line in Figure \ref{thresh_cu}).
No plateaus can be observed at higher thresholds (note the scale of the diagram in Figure \ref{thresh_cu}, where thresholds up to only 20~\% are considered) and no specific subsets of similarly entangled orbitals can be identified.
Therefore, the subset of eight weakly entangled orbitals is excluded according to the procedure in Figure \ref{flowchart}, which leaves an active space with 28 orbitals for each structure.
Since the orbitals to be excluded from the final active space are always the same (orbitals 1, 2, 9, 13, 21, 22, 28, and 31 in each diagram of Figure \ref{mutinf_cu}), the character of each orbital remains unchanged along the reaction coordinate and the degree of entanglement is stable for each orbital.
Another conclusion that can be drawn from Figure \ref{thresh_cu} is that the degree of entanglement is rather different for the bis($\mu$-oxo) and the peroxo isomers.
The peroxo motif, which is of biradical character,\cite{cramer2006} features more highly entangled orbitals as measured by the single-orbital entropy.
It shows the highest number of selected orbitals when a threshold between 5 -- 20~\% is applied, while the lowest number of orbitals is selected for the bis($\mu$-oxo) isomer for the same threshold range.

\begin{figure}[H]
 \includegraphics[width=\textwidth]{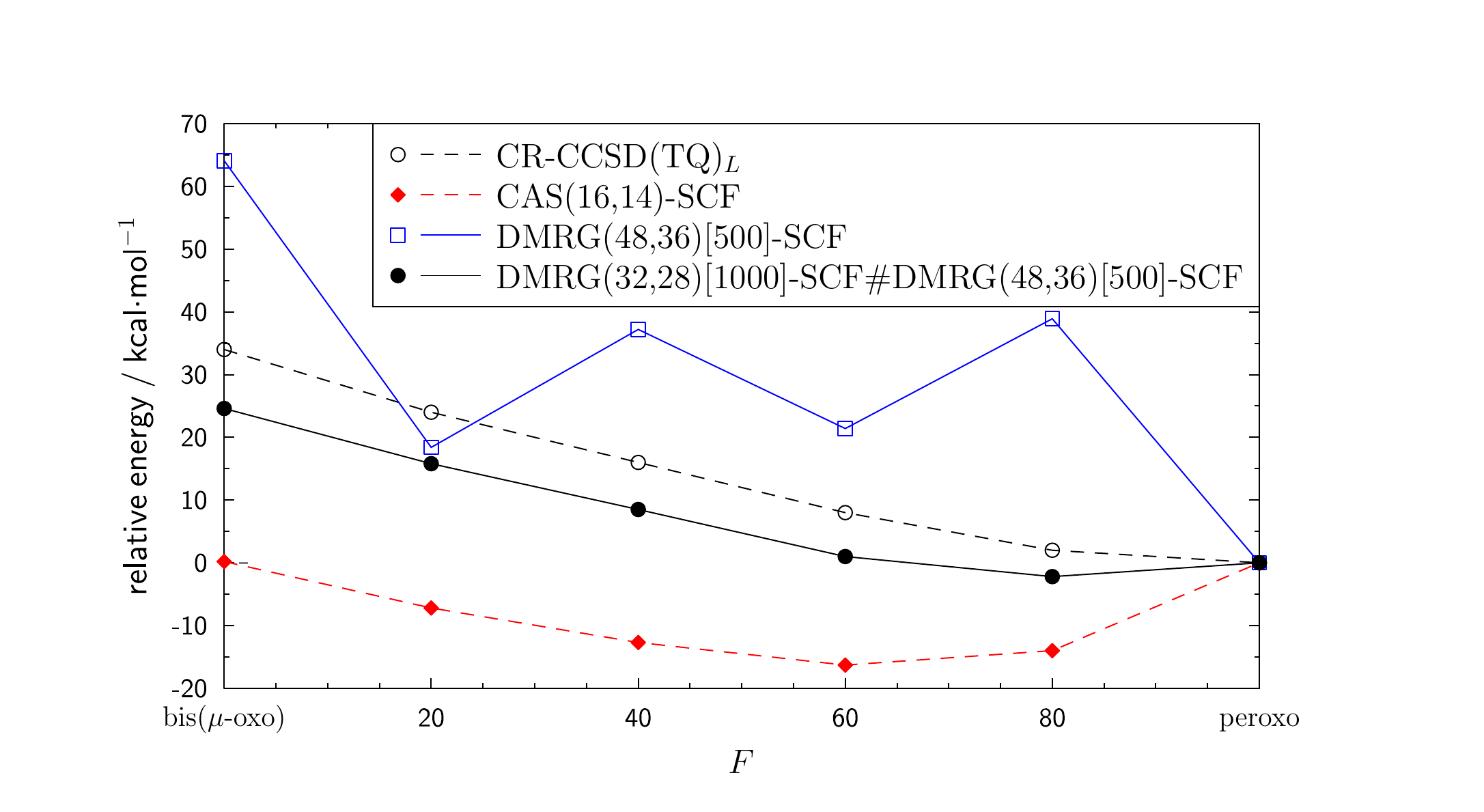}
 \caption{\small Relative energies  of six isomers of Cu$_2$O$_2^{2+}$ (the energy of the peroxo structure taken as a reference) from calculations with different methods. The CR-CCSD(TQ)$_L$ and CAS(16,14)-SCF values were taken from Ref. \citenum{gagliardi2008}.}
 \label{comp_en}
\end{figure}

After the active space selection, the orbitals were fully optimized in a DMRG(32,28)[1000]-SCF calculation.
Figure \ref{comp_en} shows the relative energies of the six structures along the isomerization coordinate with the energy of the peroxo structure taken as a reference.
We choose the coupled-cluster CR-CCSD(TQ)$_L$ energies of Ref.~\citenum{cramer2006} as benchmark results in this diagram.
As expected, the relative energies for the initial DMRG calculations scatter because convergence was not reached.
Nevertheless, these calculations allow for the selection of a suitable active space which, when further optimized (black line in Figure \ref{comp_en}), reproduces qualitatively correct relative energies. 
Clearly, the CAS selection based on the entanglement information of a partially converged calculation involving the full valence orbital space yields a wave function in the final DMRG calculations that gives qualitatively correct energies.

The energy difference between the bis($\mu$-oxo) and peroxo forms can be converged and extrapolated to a truncation error of zero as described in Section \ref{comp_sect}.
This lowers the energy gap of 24.7 kcal/mol in Figure \ref{comp_en} to 21.8~kcal/mol.
A comparison of this value with those obtained by other methods and varying sizes of the active space is presented in Table \ref{methods_comp}.

\begin{table}[H]
\caption{\small Energy difference (in kcal/mol) of the bis($\mu$-oxo) and peroxo form of Cu$_2$O$_2^{2+}$ as obtained by different CAS methods and the CR-CCSD(TQ)$_L$ best estimate energy.}
\begin{center}
\begin{tabular}{llr}
\hline
\hline
Ref. & method & $\Delta E_\mathrm{bis-per}$ \\
\hline
\citenum{gagliardi2008} & CR-CCSD(TQ)$_L$  & 34.0 \\
\citenum{cramer2006} & CAS(16,14)-SCF & 0.2\\
\citenum{gagliardi2008} & RAS(24,28)-PT2 & 28.6\\
\textit{this work}  &DMRG(32,28)[$2000,3000,4000$]\#DMRG(32,28)[$1000$]-SCF   & 21.8\\
\citenum{legeza2011} & DMRG(26,44)[1024]\#HF  & 26.8\\
\citenum{yanai2009} & DMRG(32,62)[2400]\#HF  & 35.6\\
\citenum{yanai2010} & DMRG(28,32)[2400]-SCF & 25.5 \\
\hline
\hline
\end{tabular}
\end{center}
\label{methods_comp}
\end{table}

Clearly, the CAS(16,14)-SCF energy is qualitatively wrong, while the RASPT2 result obtained for an active space size of 24 electrons in 28 orbitals is much closer to the CR-CCSD(TQ)$_L$ data.\cite{gagliardi2008} 
Unfortunately, no RASSCF values are reported in Ref.~\citenum{gagliardi2008} so that an estimate of the effect of dynamical correlation is not possible.
It is remarkable, however, that the energy difference gradually increases with the inclusion of more virtual orbitals in the active space.
The fact that the largest CAS result (35.6 kcal/mol) almost coincides with the best estimate energy difference (34.0 kcal/mol) that includes dynamical correlation is not necessarily a sign of the validity of the approach to apply increasingly large CASs. 
It might well be that even larger active spaces result in an even larger energy gap because of an imbalanced and partial treatment of the dynamical correlation.
The simple addition of more virtual orbitals is a procedure unlikely to be successful when it comes to calculations on large molecules because the ability of DMRG to also capture dynamic correlation will be limited to the growing number of orbitals in the CAS.

In this section, we determined a consistent and compact active space capable to describe the energy along an isomerization coordinate of Cu$_2$O$_2^{2+}$  from our automated approach.
This reference may then be used for the inclusion of dynamical correlation through multi-reference perturbation theory to arrive at quantitative results for this reaction.

\section{Conclusions}
We showed that suitable active orbital spaces can automatically be selected based on the orbital entanglement information of partially converged DMRG calculations with a large number of active orbitals around the Fermi level.
A tailored CAS can then be established that allows a CASSCF or DMRG calculation to cover all essential static electron correlation effects, while neglecting weakly entangled orbitals that would introduce arbitrary amounts of dynamic electron correlation into the CAS. 
Partial convergence and a small number of renormalized block states are required to quickly assess the entanglement pattern among the orbitals in the large CAS.
With this approach, we were able to reproduce active spaces for molecules identified and discussed in the literature as complicated cases.
By comparison with active spaces recommended in the literature, we concluded that the single-orbital entropy provides a more reliable selection criterion than the mutual information.
Our approach relies on the advantages of DMRG to handle large active spaces and to obtain a qualitatively converged wave function after few iterations.
Calculations with the selected active space can then be fully converged.

The selection protocol itself can be summarized as follows:
After an initial investigation of the multi-configurational character of the wave function, subsets of highly entangled orbitals are identified by automated inspection of threshold diagrams.
In cases where no subsets can be identified, orbitals with a very low single-orbital entropy are automatically excluded from the final active orbital space.
The quality of the final CAS can be assessed by a comparison of the entanglement information of the final calculation with that of the initial calculation.\\
This automated approach has the potential to overcome the extremely tedious and error prone step of CAS selection in multi-reference calculations.
If the final CAS is small, also traditional methods can be applied so that the truncation error of DMRG does not affect the final result.

In all those cases considered in this study (Sections \ref{chrome} and \ref{salen}), where previous authors applied NOON based selection criteria (or justified their choice of CAS with NOONs) we did not find a contradiction to our entanglement based active orbital selection if DMRG NOONs are considered.
An occupation number based orbital selection, however, relies on a global threshold --- usually orbitals with ONs between 0.02 and 1.98 are selected --- which is not flexible with respect to the degree of static correlation present in a molecule.
We attempt to overcome this inflexibility by coupling the determination of the orbital discarding threshold to the maximum $s_i(1)$ value of a given calculation.
This makes our approach transferable to different classes of molecules and orbital bases without the need to employ a multitude of predefined thresholds for, e.g., each type of orbital basis.

Methods that capture dynamical correlation required for accurate results rely on a balanced reference wave function and therefore on a suitable choice of CAS, which is delivered by the automated selection protocol presented here.
For "perturb-then-diagonalize" approaches our ansatz will also work as we showed that the combination of short-range DFT with DMRG has a regularizing effect on the orbital entanglement.\cite{hedegard2015}

\section*{Acknowledgements}
This work was supported by the Schweizerischer Nationalfonds (Project No. 200020\_156598).
CJS thanks the Fonds der Chemischen Industrie for a K\'ekule PhD fellowship.



\providecommand{\refin}[1]{\\ \textbf{Referenced in:} #1}


\end{document}